\documentclass{ifacconf}

\usepackage{graphicx}      % include this line if your document contains figures
\usepackage{natbib}        % required for bibliography

\usepackage{srcltx}
\usepackage{extarrows}  % to get the long equal sign etc
\usepackage{amsmath,amsfonts,amssymb,dsfont}% popular packages from the American Mathematical Society

\usepackage{fancyhdr}

\usepackage{color}
\usepackage{soul}
\newcommand{\dd}{\ensuremath{\text{d}}}

\newcommand{\eg}{\mbox{{e.g.}}}

\newcommand{\taue}{\ensuremath{\tau_\epsilon}}
\newcommand{\taus}{\ensuremath{\tau_\sigma}}
\newcommand{\taun}{\ensuremath{\tau_\nu}}

\newcommand{\kappao}{\ensuremath{\kappa(\omega)}}

\newcommand{\kappaz}{\ensuremath{\kappa_0}}
\newcommand{\kappan}{\ensuremath{\kappa_\nu}}

\newcommand{\Fouriertransform}[1]{\ensuremath{\mathcal F}\left\{{#1}\right\}}

\newcommand{\Fourierbackfourth}{\ensuremath{\xleftrightarrow{\ \textstyle\mathcal F\ }}}

\newcommand{\C}{\ensuremath{\mathbb{C}}}

\newcommand{\MittagLeffler}[3]{\ensuremath{E_{#1,#2} \big(#3\big)}}
\newcommand{\MittagLefflerAlphaBetaT}{\MittagLeffler{a}{b}{t}}

	\newcommand{\maybehighlight}[1]{{#1}} % DO NOT HIGHLIGHT
	
\begin{document}
\begin{frontmatter}

\title{A Fractional Acoustic Wave Equation from Multiple Relaxation Loss and Conservation Laws\thanksref{footnoteinfo}}  

\thanks[footnoteinfo]{This research was partly supported by the ``High Resolution Imaging and Beamforming'' project of the Norwegian Research Council.}

\author{Sven Peter N\"asholm and} 
\author{Sverre Holm} 
%\author[second]{Ralph Sinkus}

\address{Department of Informatics, University of Oslo, P.\hspace{.5pt}O.~Box 1080, NO--0316 Oslo, Norway (e-mail: svenpn@i{f}i.uio.no).}
%\address[second]{Department of Radiology, CRB3, H\^{o}pital Beaujon (U773), INSERM, Clichy, France}

\begin{abstract}                % Abstract of not more than 250 words.
This work concerns causal acoustical wave equations which imply frequency power-law attenuation. A connection between the five-parameter fractional Zener wave equation, which is derived from a fractional stress-strain relation plus conservations of mass and momentum, and the physically well established multiple relaxation framework is developed. %
It is shown that for a certain continuous distribution of relaxation mechanisms, the two descriptions are equal. %This work hence provides a physically based motivation for use of fractional wave equations in acoustic modeling.
\end{abstract}


\end{frontmatter}
%===============================================================================

\section{Introduction}

The multiple relaxation mechanism framework of \cite{Nachman1990} is widely considered as adequate for acoustic wave %
modeling in lossy complex media like those encountered in medical ultrasound. % 
It relies on thermodynamics and first principles of acoustical physics. The corresponding wave equation for $N$ relaxation mechanisms is a causal partial differential equation with its highest time derivative order $N+2$. We denote this the {Nachman--Smith--Waag (NSW) model}. %

Attenuation in complex media often follows a power law: $\alpha_k(\omega) \propto \omega^\eta$, with $\eta\in[0,2]$ (\cite{Szabo00}). The range where experiments indicate this may cover many frequency decades. In order to make the NSW model attenuation adequately follow $\omega^\eta$, either the valid frequency band must be narrow, or the number of assumed mechanisms $N$ must be large thus inferring a partial differential equation of very high order. %

Another way to derive a lossy wave equation is to combine the principles of mass and momentum conservation  with some stress--strain relation. %
This constitutive relation may include fractional time-derivatives, exemplified by the fractional Zener model  by \cite{Holm2011}. % 
The resulting wave equation is causal and the corresponding attenuation follows power laws within wide frequency bands. %

The purpose of the present work is to demonstrate the link between the NSW and the fractional Zener models via a continuum of relaxation mechanisms. %
Relevant parts of \cite{Nasholm2011} are reviewed and reformulated. In addition a more general deduction is provided where the fractional Zener model parameters $\alpha$ and $\beta$ are not necessarily equal. %
We aim to encourage the acoustical community to more frequently adopt fractional calculus descriptions for wave modeling in complex media. %more achieve a joint description where the advantages of the fractional Zener model are combined with those of the NSW model. %

\section{Theory}

%\subsection{Conservation of Mass, Conservation of Momentum, and Generalized Compressibility}
\subsection{Conservation laws and Generalized Compressibility}
The linearized conservation of mass corresponds to the strain being defined by
\begin{align}
	\epsilon(t) = \nabla u(x,t), \Fourierbackfourth \epsilon(\omega) = -i k\:u(k, \omega),
	\label{Eq:strain}
\end{align}
where $u$ is the displacement and the symbol $\mathcal F$ denotes transformation into the spatio-temporal frequency domain where $\omega$ is the angular frequency and $k$ the wavenumber. %

The linearized conservation of momentum is expressed as
\begin{align}
	\nabla \sigma(t) = \rho_0 \frac{\partial^2u(x,t)}{\partial t^2}, \Fourierbackfourth -i k \sigma(\omega) = \rho_0 (i \omega)^2 u(k,\omega),
	\label{Eq:Newton}
\end{align}
where $\rho_0$ is the steady-state mass density and $\sigma$ denotes the stress, which in this context corresponds to the negative of the pressure. %

The frequency-domain generalized compressibility is defined as the ratio between strain and stress: $\kappa(\omega) \triangleq \epsilon(\omega)/\sigma(\omega)$, therefore being related to the constitutive stress--strain relation. %
Combining this definition with the conservation laws \eqref{Eq:strain} and \eqref{Eq:Newton} gives 
\begin{align}
	k^2&(\omega) = \omega^2\rho_0 \kappa(\omega) \label{eq:dispersion_relation}\\
	&\Fourierbackfourth %
	\nabla^2u(x,t) - \dfrac{\dd^2}{\dd t^2} \left[\kappa(t) \underset{t}{*} u(x,t)\right] = 0.
	\label{eq:general_dispersion_relation}
\end{align}
Under circumstances where the linearized conservations of mass and momentum are valid, the wave equation is thus completely determined by the generalized compressibility. %

\maybehighlight{The generalized compressibility $\kappa(\omega)$ as given above is sometimes (\eg{} in viscoelasticity) called complex compliance $J^*(\omega)=1/G^*(\omega)$, where $G^*(\omega)$ is the complex modulus.} %

\subsection{A Continuum of NSW Relaxation Processes}
The NSW model of multiple discrete relaxation processes results in the generalized compressibility 
\begin{align}
	\kappao = \kappaz - i\omega \sum_{\nu=1}^N \dfrac{\kappan \taun}{1 + i\omega\taun},
	\label{eq:Nachman_kappa_omega}
\end{align}
where the mechanisms $\nu=1\ldots N$, have the relaxation times $\tau_1,\ldots,\tau_N$ and the compressibility contributions $\kappa_1, \ldots, \kappa_N$ (\cite{Nachman1990}). %

Following \cite{Nasholm2011}, a representation of \eqref{eq:Nachman_kappa_omega} when considering a continuum of relaxation mechanisms distributed in the frequency band $\Omega \in[\Omega_1,\Omega_2]$ with the compressibility contributions described by the distribution $\kappa_\nu(\Omega)$ becomes
\begin{align}
	 \kappa_\text{N}(\omega) \triangleq \kappaz - i\omega\int_{\Omega_1}^{\Omega_2} \dfrac{ \kappan(\Omega) }{\Omega + i\omega}\, \dd \Omega.
	\label{eq:Nachman_kappa_omega_integral_zenertry}
\end{align}
Letting the limits of the integral go between $\Omega_1=0$ and $\Omega_2=\infty$, and instead incorporating any possible relaxation distribution bandwidth limitation into $\kappa_\nu(\Omega)$, the integral above is a Stieltjes transform. Applying the Laplace transform relation
\begin{align}
	\mathcal L^{-1}_\Omega\left\{\dfrac 1 {\Omega+i\omega}\right\}(t) = e^{-i\omega t},
\end{align}
the generalized compressibility \eqref{eq:Nachman_kappa_omega_integral_zenertry} becomes %
\begin{align}
	\kappa_\text{N}(\omega)
	&= \kappa_0 -i\omega \int_0^\infty \kappan(\Omega) \int_0^\infty e^{-\Omega t} e^{-i\omega t} \dd t\; \dd \Omega\notag\\
	\ &=  \kappa_0 -i\omega \mathcal F_t \Big\{H(t) \mathcal L_\Omega \left\{ \kappan(\Omega)  \right\}\!\! (t) \Big\}(\omega).
	\label{eq:nachman_fourier_laplace}
\end{align}

\subsection{The Fractional Zener Wave Equation}
The five-parameter fractional Zener stress--strain constitutive relation is experimentally shown to be valid for a wide range of complex media, see the references in \cite{Holm2011}. As given by \cite{Bagley83A}, it may be expressed as %
\begin{align}
	\sigma(t) +\tau_{\epsilon}^{\beta} \frac{\partial^{\beta}\sigma(t)}{\partial t^{\beta}}  = E_0 \left[\epsilon(t) +\tau_{\sigma}^{\alpha} \frac{\partial^{\alpha}\epsilon(t)}{\partial t^{\alpha}}\right].
	\label{Eq:gZener}
\end{align} 
From this relation, the frequency-domain fractional Zener compressibility is obtained through the ratio $\epsilon(\omega)/\sigma(\omega)$:
\begin{align}
	\kappa_\text{Z}(\omega) & \triangleq %
		\kappaz \frac{1 + (\tau_{\epsilon}i \omega)^{\beta}}{1+ (\tau_{\sigma}i \omega)^{\alpha}} \notag\\
	&= %
		\kappaz -i\omega%\kappaz(1-\tau_\epsilon^\beta/\tau_\sigma^\alpha)  \frac{(i \omega)^{\beta-1}}{\tau_\sigma^{-\alpha}+ (i \omega)^{\alpha}}. %
	\kappa_0 \dfrac{ (i\omega)^{\alpha-1} - (\tau_\epsilon^\beta/\tau_\sigma^\alpha)(i\omega)^{\beta-1}}{ \tau_\sigma^{-\alpha} + (i\omega)^\alpha}
\label{Eq:fZener_compressibility}
\end{align}
Due to thermodynamic constraints, $\beta$ is restricted to be smaller than or equal to $\alpha$ (\cite{Glockle1991}). %

Insertion of the generalized compressibility \eqref{Eq:fZener_compressibility} into the dispersion relation \eqref{eq:general_dispersion_relation}, generates the time-domain fractional Zener wave equation \maybehighlight{(}\cite{Holm2011}\maybehighlight{)}
\begin{align}
	{\nabla^2 u -\dfrac 1{c_0^2}\frac{\partial^2 u}{\partial t^2} + \taus^\alpha \dfrac{\partial^\alpha}{\partial t^\alpha}\nabla^2 u	- \dfrac {\taue^\beta}{c_0^2} \dfrac{\partial^{\beta+2} u}{\partial t^{\beta+2}} = 0.}
	\label{Eq:wave_equation_zener}
\end{align}

\subsection{Connecting the NSW and the Fractional Zener Models}
Provided that the conservations of mass \eqref{Eq:strain} and momentum \eqref{Eq:Newton} are valid, and provided that the NSW generalized compressibility $\kappa_\text{N}(\omega)$ of \eqref{eq:nachman_fourier_laplace} 
is equal to the fractional Zener generalized compressibility $\kappa_\text{Z}(\omega)$ of \eqref{Eq:fZener_compressibility}, %
the dispersion relations from \eqref{eq:dispersion_relation} are also equal. Because the dispersion relation is a spatio-temporal Fourier representation of the wave equation, %
$\kappa_\text{N}(\omega) = \kappa_\text{Z}(\omega)$ thus implies that the NSW wave equation becomes equal to the fractional Zener wave equation \eqref{Eq:wave_equation_zener}. %
Direct comparison of $\kappa_\text{N}(\omega)$ in \eqref{eq:nachman_fourier_laplace} to $\kappa_\text{Z}(\omega)$ in \eqref{Eq:fZener_compressibility}, tells that they are equal in case the following is true:
\begin{align}
	\mathcal F_t \Big\{H(t) \mathcal L_\Omega & \left\{ \kappan(\Omega)  \right\}\!\! (t) \Big\}(\omega)	
	\notag\\
		&=
	\kappa_0 \dfrac{ (i\omega)^{\alpha-1} - (\tau_\epsilon^\beta/\tau_\sigma^\alpha)(i\omega)^{\alpha-(\alpha-\beta+1)}}{ \tau_\sigma^{-\alpha} + (i\omega)^\alpha}.
	\label{eq:compressibilities_are_equal}
\end{align}

First we choose to study the case $\alpha=\beta$, which was also treated in \cite{Nasholm2011}. Inverse Fourier transformation of both sides of \eqref{eq:compressibilities_are_equal}, then gives
\begin{align}
	H(t) &\mathcal L_\Omega \left\{ \kappan(\Omega)  \right\}\! (t)	\notag\\
		&=
	\kappa_0 (1-\tau_\epsilon^\alpha/\tau_\sigma^\alpha) \mathcal F_\omega^{-1} \left\{ \dfrac{ (i\omega)^{\alpha-1}}{ \tau_\sigma^{-\alpha} + (i\omega)^\alpha}\right\}\!(t)\notag\\
		&=
		\kappa_0 (1-\tau_\epsilon^\alpha/\tau_\sigma^\alpha) H(t) E_{\alpha,1} \left(-(t/\tau_\sigma)^\alpha \right),
	\label{eq:compressibilities_are_equal_alphaisbeta}
\end{align}
where $E_{a,b}( \cdot )$ is the Mittag-Leffler function (see Appendix \ref{section:ML_appendix}), and $H(t)$ is the Heaviside step function. The Fourier transform relation used in the last step above is given in \eqref{eq:mittagleffler_fourier_transform}. %
Moreover, the inverse Laplace transform relation of \eqref{eq:mittagleffler_integral_relation}, Eq.~\eqref{eq:compressibilities_are_equal_alphaisbeta} hence gives
\begin{align}
	\kappan(\Omega) &=
	\kappa_0 (1-\tau_\epsilon^\alpha/\tau_\sigma^\alpha) f_{\alpha,1}\left(\Omega, \tau_\sigma^{-\alpha}\right)
	\notag\\
	& = \dfrac{1}{\pi} \dfrac{\kappaz(\taus^{\alpha}- \taue^\alpha)\Omega^{\alpha-1} \sin (\alpha\pi ) }{ (\taus\Omega)^{2\alpha} + 2(\taus\Omega)^\alpha \cos(\alpha\pi) + 1}
	\triangleq \kappa_{\nu\text{ML}}(\Omega) 
	\label{eq:distribution_for_alphaisbeta}
\end{align}
where $f_{\alpha,1}(\Omega,a)$ was inserted from \eqref{eq:f_alpha_beta_distribution}.

For the more general case $\beta \leq \alpha$, inverse Fourier transform on both sides of \eqref{eq:compressibilities_are_equal} instead gives%
\begin{align}
	H(t) &\mathcal L_\Omega \left\{ \kappan(\Omega)  \right\}\! (t)	=\notag\\
		&\quad
	\kappa_0 \mathcal F_\omega^{-1} \left\{ \dfrac{ (i\omega)^{\alpha-1}}{ \tau_\sigma^{-\alpha} + (i\omega)^\alpha}\right\}\!(t) \notag\\
	&-\kappa_0 (\tau_\epsilon^\beta/\tau_\sigma^\alpha) \mathcal F_\omega^{-1} \left\{ \dfrac{ (i\omega)^{\alpha-(\alpha-\beta+1)}}{ \tau_\sigma^{-\alpha} + (i\omega)^\alpha}\right\}\!(t).
	\label{eq:compressibilities_are_equal_alphaisNOTbeta}
\end{align}
Proceeding in a similar manner as for the $\alpha=\beta$ case then gives the distribution
\begin{align}
	&\kappan(\Omega) =
	\kappa_0 f_{\alpha,1}\left(\Omega, \tau_\sigma^{-\alpha}\right) 
		-\kappa_0(\tau_\epsilon^\beta/\tau_\sigma^\alpha) f_{\alpha,\alpha -\beta+1}\left(\Omega, \tau_\sigma^{-\alpha}\right)
	\notag\\
	& = \dfrac{\kappa_0\tau_\sigma^\alpha}{\pi}  \cdot %
		\dfrac{ \Omega^{\alpha-1} \sin (\alpha\pi )}%
		{ (\taus\Omega)^{2\alpha} + 2(\taus\Omega)^\alpha \cos(\alpha\pi) + 1}\notag\\
		&\quad-\dfrac{\kappa_0\tau_\epsilon^\beta }{\pi} \cdot \dfrac{\Omega^{\beta-1}\sin(\beta\pi) - \tau_\sigma^\alpha\Omega^{\alpha+\beta-1} \sin ((\alpha-\beta)\pi)}%
		{ (\taus\Omega)^{2\alpha} + 2(\taus\Omega)^\alpha \cos(\alpha\pi) + 1}
	\notag\\
	&\triangleq \kappa_{\nu\text{ML}}'(\Omega).
	\label{eq:distribution_for_alphaisNOTbeta}
\end{align}

We have thus shown that the fractional Zener wave equation \eqref{Eq:wave_equation_zener} may be obtained within the Nachman--Smith--Waag framework of multiple relaxation (\cite{Nachman1990}), when assuming a continuum of relaxation mechanisms with the compressibility contribution as given by the distribution $\kappa_{\nu\text{ML}}'(\Omega)$ of \eqref{eq:distribution_for_alphaisNOTbeta}.

\maybehighlight{In viscoelasticity, a relaxation time spectrum (below denoted $\tilde H(\tau)$) related to $\kappa_\nu(\Omega)$ is commonly studied (see \eg{} }\cite{Glockle1991}\maybehighlight{ and references therein). It is related to the complex modulus through}
\begin{align}
	G(t) = G_\infty + \int_{-\infty}^\infty  \tilde H(\tau) e^{-t/\tau} \dd \ln \tau.
\end{align}
\maybehighlight{It may be shown that for the 5-parameter fractional Zener model, when setting $\Omega=\tau^{-1}$, the $\tau$-dependency of $\tilde H(\tau)$ differs by a factor $\tau$ to $\kappa_{\nu\text{ML}}(\Omega)$ of }\eqref{eq:distribution_for_alphaisNOTbeta}\maybehighlight{. %
Figs.~5 and 6 of }\cite{Glockle1991}\maybehighlight{ illustrate that $\alpha=\beta$ gives symmetric $\tilde H(\tau)$, while $\alpha\neq\beta$ breaks the symmetry, most significantly far from the peak region. %
Such relaxation spectra are experimentally observed for complex media, \eg{} natural rubber. }

\section{Attenuation and phase velocity examples}
The conventional decomposition of the frequency-dependent wavenumber into its real and imaginary parts, 
gives the phase velocity $c_p(\omega) = \omega/\Re\left\{k\right\}$ and attenuation $\alpha_k(\omega) = -\Im\left\{k\right\}$. %

In general, the attenuation and the phase velocity are thus given from the dispersion relation \eqref{eq:dispersion_relation} as
\begin{align}
\begin{array}{l}
	\alpha_k(\omega) = -\Im\left\{k\right\} = -\omega\sqrt{\rho_0}\Im\left\{\sqrt{\kappa(\omega)}\right\}	\quad\text{and}\\
	c_p(\omega) = \omega/\Re\left\{k\right\} = {\rho_0^{-1/2}}/\Re\left\{\sqrt{\kappa(\omega)}\right\}.
	\label{eq:atten_and_soundspeed_from_kappa}
	\end{array}
\end{align}
For the fractional Zener wave equation, this results in three distinct regions with attenuation power-laws (\cite{Nasholm2011}): $\alpha_k \propto \omega^{1+\alpha}$ in a low-frequency regime, $\alpha_k \propto \omega^{1-\alpha/2}$ in an intermediate frequency regime, and $\alpha_k \propto \omega^{1-\alpha}$ in a high-frequency regime.

In the following, the fractional Zener phase velocities and attenuations are further investigated numerically for the $\alpha=\beta$ case in a similar manner as in \cite{Nasholm2011}. %
This is done explicitly by insertion of $\kappa(\omega) = \kappa_\text{N}(\omega)$ into \eqref{eq:atten_and_soundspeed_from_kappa}. %
The results from such calculations are compared to what is found by insertion of the distribution $\kappa_{\nu\text{ML}}(\Omega)$ of \eqref{eq:distribution_for_alphaisbeta} into the NSW generalized compressibility integral formula \eqref{eq:Nachman_kappa_omega_integral_zenertry}. This generalized compressibility is finally applied to \eqref{eq:atten_and_soundspeed_from_kappa}, from which $\alpha_k(\omega)$ and $c_p(\omega)$ are found. %

We use the latter calculation method to explore the effect of letting the continuum of relaxation mechanisms populate only a bounded frequency interval, rather than the entire $\Omega\in[0,\infty]$ region. %

\begin{figure*}[th!]
	\begin{center}
		\includegraphics[width=\textwidth]{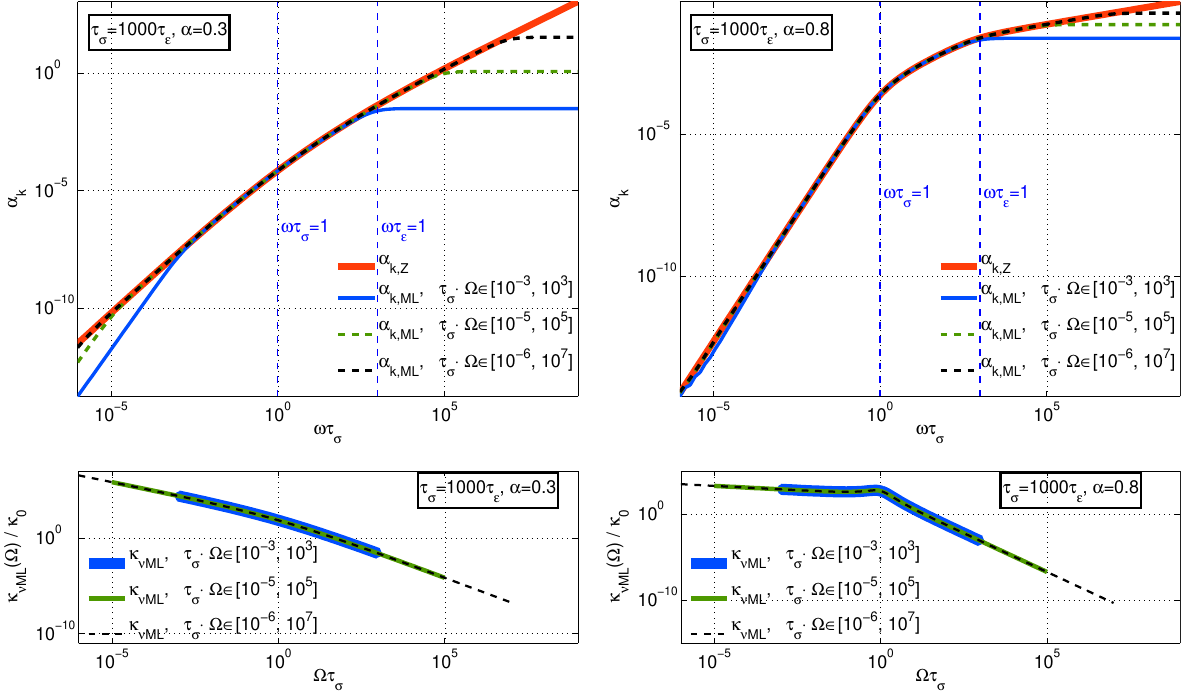}
	\end{center}

	\caption{\label{fig:attenuation}
		Top panes: frequency-dependent attenuation for $\taus=1000\taue$ with the fractional derivative orders $\alpha=0.3$ (top-left pane), and $\alpha=0.8$ (top-right pane). %
		The attenuation curves display both explicit calculations from the fractional Zener model, and calculations using the distribution $\kappa_{\nu\text{ML}}(\Omega)$ in the NSW model. Different choices of integration limits for $\Omega_1$ and $\Omega_2$ in \eqref{eq:Nachman_kappa_omega_integral_zenertry} are exploited, as displayed in the legends. %
		The horizontal axis represents normalized frequency. For visualization convenience, each attenuation curve is normalized to $\alpha_k=1$ at $\omega\taus=1$. %\\%
	Bottom panes: the corresponding normalized effective compressibilities $\kappa_{\nu\text{ML}}(\Omega)$ of the continuum of relaxation processes as a function of normalized relaxation frequency $\Omega\cdot\taus$ for $\alpha=0.3$ (bottom-left pane), and $\alpha=0.8$ (bottom-right pane). %
	The $\Omega$ integration limits are given in the legends.}
\end{figure*}

\begin{figure*}[th!]
	\begin{center}
		%\insfigdoublecolumn{soundspeed_0_3} \insfigdoublecolumn{soundspeed_0_8}
		\includegraphics[width=\textwidth]{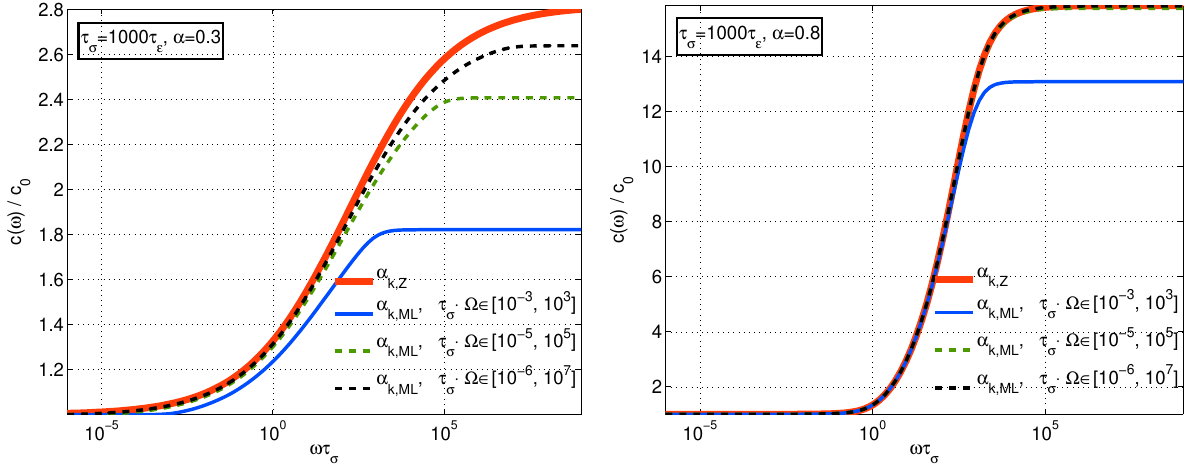}
	\end{center}
	\caption{\label{fig:soundspeed}Frequency-dependent phase velocity for $\taus=1000\taue$ with the fractional derivative orders $\alpha=0.3$ (left pane), and $\alpha=0.8$ (right pane). The curves display $c_p(\omega)$ as predicted by the fractional Zener model, as well as predicted by 
	using the distribution $\kappa_{\nu\text{ML}}(\Omega)$ in the NSW model. Different choices of integration limits for $\Omega_1$ and $\Omega_2$ in \eqref{eq:Nachman_kappa_omega_integral_zenertry} are exploited, as displayed in the legends. %
 The horizontal axis represents normalized frequency.}
\end{figure*}

Figure \ref{fig:attenuation} compares attenuation curves and shows the distributions $\kappa_\nu(\Omega)$, while Fig.~\ref{fig:soundspeed} displays the corresponding frequency-dependent phase velocity. %
Note the high-frequency asymptote of $\kappa_\nu(\Omega)$, which has the fractal property of being proportional to $\Omega^{-\alpha-1}$. 
The integral over $\Omega \in[\Omega_1,\Omega_2]$ in the calculation of $\kappa_\text{N}(\omega)$ from \eqref{eq:Nachman_kappa_omega_integral_zenertry} is evaluated numerically using the recursive adaptive Simpson quadrature method. %


\section{Conclusion}
This work shows analytically that the lossy fractional Zener wave equation \eqref{Eq:gZener} (\cite{Holm2011}) may be attained within the NSW multiple relaxation loss framework (\cite{Nachman1990}), given that a continuum of relaxation processes are weighted appropriately following $\kappa_{\nu\text{ML}}'(\Omega)$ as described in \eqref{eq:distribution_for_alphaisNOTbeta}. %
The result may be seen as a generalization of what was presented in \cite{Nasholm2011} as the developments here also cover the case of the fractional Zener constitutive relation not having equal derivative orders $\alpha$ and $\beta$.

The advantages of the fractional Zener model is that it fits measurements well and that it is characterized by a small number of parameters, while the NSW model is more intuitive as it does not comprise fractional derivatives. It is also better rooted in fundamental physics. %

%\begin{ack}
%This research was partly supported by the ``High Resolution Imaging and Beamforming'' project of the Norwegian Research Council.
%\end{ack}

%\bibliography{Memory-constitutive}             % bib file to produce the bibliography

                                                     % with bibtex (preferred)

\appendix
\section{The Mittag-Leffler function} % 
{\label{section:ML_appendix}
% =========================================================
%
%
%
\subsection{Definition and Fourier Transform Relation}
The one-parameter Mittag-Leffler function was introduced in \cite{Mittag-Leffler1903}. A two-parameter analogy was presented in \cite{Wiman1905}, which may be written as
\begin{align}
	\MittagLefflerAlphaBetaT \triangleq \sum_{n=0}^\infty \dfrac{t^n}{\Gamma(a n+ b )},
	\label{eq:mittagleffler_definition}
\end{align}
where $\Gamma$ is the Euler Gamma function and the parameters are commonly restricted to $\{a, b \} \in \C,\ \Re{\{a, b \}}>0$, and $t\in\C$. %
%The one-parameter Mittag-Leffler function $E_a(t)$ equals the two-parameter $E_{ a , b }(t)$ of \eqref{eq:mittagleffler_definition} with $ b =1$. 
See~\cite{Haubold2011} for a comprehensive review of Mittag-Leffler function properties. 

A useful Fourier transform pair involving the Mittag-Leffler function is (\cite{Podlubny1999chapter1-2})
\begin{align} 
	\Fouriertransform{H(t)\: t^{ b -1}\MittagLeffler{ a }{ b }{-A t^ a }}(\omega) = & \dfrac{(i\omega)^{ a - b }}{A+ (i\omega)^ a }.%,
	\label{eq:mittagleffler_fourier_transform}
\end{align}
%
\iffalse
	which is equivalent to
	%
	\begin{align} 
		\Fouriertransform{\dfrac{\dd}{\dd t}\left[ H(t)\: t^{ b -1}\MittagLeffler{ a }{ b }{-A t^ a }\right]}(\omega) = & \dfrac{(i\omega)^{ a - b  +1}}{A+ (i\omega)^ a }.
		%
		\label{eq:laplace_to_ML-related}
	\end{align}
	%

	Another useful relation which follows from the Mittag-Leffler function definition \eqref{eq:mittagleffler_definition} is
	%
	\begin{align}
		t^{ a -1}\MittagLeffler{ a }{ a }{-A t^ a }= -{A}^{-1} \dfrac{\dd}{\dd t} \MittagLeffler{ a }{1}{-A t^ a }.
		%
		\label{eq:mittagleffler_derivative_relation}
	\end{align}
\fi

\subsection{\label{section:ML_appendix_integral_representation}Laplace Transform Integral Representation}
The function $t^{b-1}\MittagLeffler{ a }{ b }{{-A}t^a}$ may for $0< a \leq 1$ be written on an integral form (\cite{Djrbashian1966,Djrbashian1993chapter1}):
\begin{align}
	t^{ b -1}\MittagLeffler{ a }{ b }{-A t^ a } = \int_0^\infty e^{-\Omega t} f_{ a , b }(\Omega, A)\: \dd \Omega,
	\label{eq:mittagleffler_integral_relation}
\end{align}
where 
\begin{align}
	f_{ a , b }(\Omega,A) = \dfrac{\Omega^{ a - b }}{\pi} \dfrac{A \sin [( b - a )\pi ] + \Omega^ a  \sin( b \pi)}{ \Omega^{2 a } + 2 A \Omega^ a  \cos( a \pi) + A^2}.
		\label{eq:f_alpha_beta_distribution}
\end{align}
%
\iffalse
	For $ b =1$, 
	\begin{align}
		f_{ a ,1}(\Omega,A) = \dfrac{1}{\pi} \dfrac{A\Omega^{ a -1} \sin ( a \pi ) }{ \Omega^{2 a } + 2 A \Omega^ a  \cos( a \pi) + A^2},
		%
		\label{eq:f_alpha_1_distribution}
	\end{align}
	%
	which has a finite integral from $0$ to $\infty$ given by
	%
	\begin{align}
		\int_0^\infty f_{ a ,1}(\Omega,A)\: \dd\Omega = \MittagLeffler{ a }{1}{0} = 1.
		\label{eq:f_alpha_1_distribution_integral}
	\end{align}
	%

	The function $f_{ a ,1}(1,\Omega)$ may be considered as a random variable density corresponding to a Mittag-Leffler distribution function $F_ a (x)=1-E_{ a ,1}(-x^ a )$ (\cite{Pillai1990}).
\fi

\iffalse
	A noteworthy observation regarding the integral \eqref{eq:f_alpha_1_distribution_integral} between the limits $\Omega=\Omega_1$ and $\Omega_2$ is that using the variable change $u=\Omega^ a $, it may instead be written
	%
	\begin{align}
		&\dfrac{A\sin ( a \pi ) }{\pi} \int_{\Omega_1}^{\Omega_2} \dfrac{\Omega^{ a -1} }{ \Omega^{2 a } + 2 A \Omega^ a  \cos( a \pi) + A^2}\: \dd\Omega \notag\\
		%
		&=\dfrac{A\sin ( a \pi ) }{ a \pi} \int_{\Omega_1^ a }^{\Omega_2^ a } \dfrac{ 1 }{ u^{2} + 2 A u \cos( a \pi) + A^2}\: \dd u.
		\label{eq:f_alpha_1_distribution_integral_varchange}
	\end{align}
\fi
	
} % end of appendix. Inserted between {} for \label{} command reasons (is this necessary?)

\end{document}